A Survey of the Latin American High Energy Physics community on the future flagship project at CERN for the ESPP Update

Reina Camacho (LPNHE/CNRS, France) [0000-0002-9192-8028], Melissa Cruz (Universidad Nacional Autónoma de Honduras) [0000-0003-2607-131X], Salvatore Mele (CERN) [0000-0003-0762-2235], Fernando Monticelli (Universidad Nacional de La Plata, Argentina) [0000-0002-6974-1443], Martijn Mulders (CERN) [0000-0001-7432-6634], Rogerio Rosenfeld (Universidade Estadual Paulista, Brazil) [0000-0001-9427-9812], Carlos Sandoval (Universidad Nacional de Colombia) [0000-0003-1038-723X] and Maria Elena Tejeda-Yeomans (Universidad de Colima, México) [0000-0002-4938-7563]
On behalf of LAA-HECAP

May 2025

Summary
This document collects input from Latin America as a contribution to the Update of the European Strategy for Particle Physics. It emerges from a survey of members of the Latin American Association for High Energy, Cosmology and Astroparticle Physics (LAA-HECAP) that collected data in February and a subsequent town-hall meeting, inspired by the ECFA guidelines for national communities. This contribution first reviews the Latin American participation at CERN, provides background on LAA-HECAP, and then presents the survey methodology and its results. Some conclusions are drawn based on the results of the survey.

1. **Latin American participation at CERN**

Individual researchers from Latin America have worked at CERN since the early years of the laboratory, and are today a vibrant community well integrated in CERN's experimental programme [1]. As of September 2024, CERN counted over 300 'users' from 40 institutes in 10 Latin American countries, about 2.5% of the total - see table below. 'Users' are students, researchers or technologists visiting the laboratory for at least 5% of their working time over a 5-year window. In the case of Latin America, this is a somewhat imperfect measure that underestimates the engagement in the region, as both the effect of the COVID-19 restrictions on international travel, and the geographical distance, compounded with relatively high travel costs, results in local communities that are between 3 and 5 times larger. Recent improvement in remote participation, also brought about by the pandemic, allows more (under-)graduate students and engineers to participate in activities remotely, albeit imperfectly. Within these caveats, it is interesting to compare the count of 'Users' affiliated to institutes in Latin America with the overall count of CERN 'Users', as depicted in Figure 1. Also thanks to external financing schemes made possible by funding by the European Commission, the number of Latin American users grew rapidly in the decade from 2005 to 2015, with a rate outpacing overall growth. The recovery from the drops brought about by the pandemic years, has equally advanced at a faster clip.

---

**Argentina:** *Universidad de Buenos Aires*, *Universidad Nacional de la Plata*.
**Brazil:** *Centro Brasileiro de Pesquisas Físicas*, *Universidade Federal do Rio de Janeiro*, *Universidade do Estado do Rio de Janeiro*, *Universidade Estadual de São Paulo*, *Universidade Estadual de Campinas*, *Universidade de São Paulo*, *Pontifícia Universidade Católica do Rio de Janeiro*, *Universidade Federal da Bahia*, *Universidade Federal do ABC*, *Universidade Federal de Juiz de Fora*, *Universidade Federal do Rio Grande do Sul*
**Chile:** *Universidad Técnica Federico Santa María*, *Pontificia Universidad Católica de Chile*, *Universidad Andrés Bello*, *Universidad de La Serena*, *Universidad de Tarapacá*, *Comisión Chilena de Energía Nuclear*
**Colombia** *Universidad de Antioquia*, *Universidad Nacional*, *Universidad Antonio Nariño*, *Universidad de los Andes* and *Universidad del Valle*
**Costa Rica:** *Universidad de Costa Rica*, *Instituto Tecnológico de Costa Rica*, *Universidad Nacional*, *Universidad Estatal a Distancia*, *Universidad Técnica Nacional*
**Cuba:** *Centro de Aplicaciones Tecnológicas y Desarrollo Nuclear*
**Ecuador:** *Universidad San Francisco de Quito*, *Escuela Politécnica Nacional*
**Mexico:** *Benemérita Universidad Autónoma de Puebla*, *Centro de Investigación y de Estudios Avanzados*, *Universidad Nacional Autónoma*, *Universidad Iberoamericana*, *Universidad Autónoma de San Luis Potosí*, *Universidad Autónoma de Sinaloa*, *Universidad de Sonora*
**Peru:** *Pontificia Universidad Católica de Perú*

---

Table 1: Institutions from Latin America that have users at CERN (from [1]).

A different picture emerges from an analysis of the nationality of all CERN 'Users': the number of Latin American nationals climbs to about 4% of the total (while only 2.5% is affiliated to institutes in the region, and not necessarily in their country of origin). This bears witness both to the talent of the region, and the incentives to seek positions outside the region at different moments of an individual's career [1]. Specifically, while 30% of 12,726 CERN 'Users' are affiliated in a country other than their nationality, this number climbs to 47% for Latin American nationals (of those, 53% are affiliated to institutes in Europe, 39% in the U.S. and Canada, 5% elsewhere in Latin America, 4% in the rest of the world). It should be also noted that 11% of 'Users' in Latin American institutes are not Latin American nationals.

These participation of Latin American institutes in CERN's programmes are underpinned by the institutional participation of Latin American countries at CERN. Those are underpinned by International Cooperation Agreements that the governments of 12 countries have signed with CERN: Argentina, Brazil, Bolivia, Chile, Colombia, Costa Rica, Ecuador, Honduras, Mexico, Paraguay, Peru, and Uruguay. Brazil became a CERN Associate Member State, the first in the Americas, in 2024, while Chile is well advanced in this process.

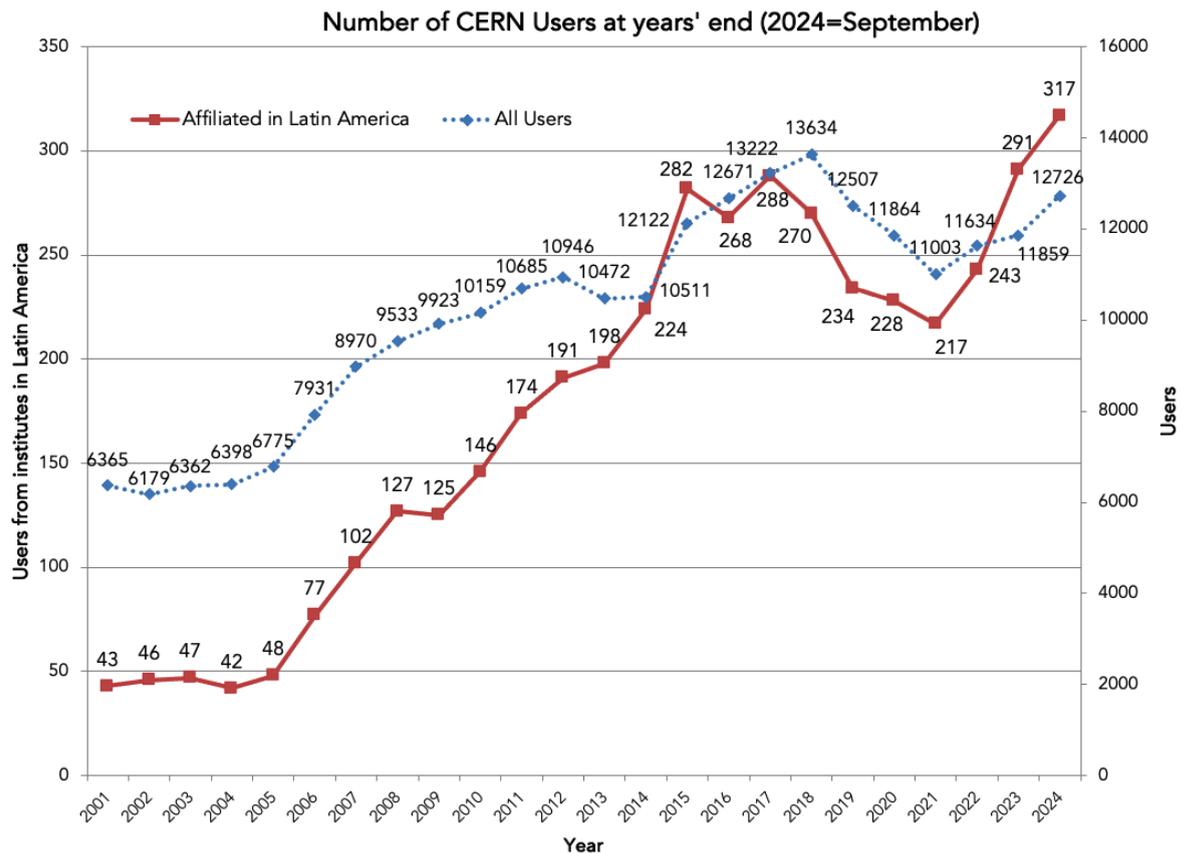

Figure 1: Counts of CERN 'Users' (Students, researchers or engineers visiting the laboratory for at least 5% of their working time over a 5-year window) as a function of time, for all institutes participating in CERN's physics programmes, and those established in Latin America (from [1]).

Looking ahead, researchers from 13 institutes in Latin America (from Argentina, Brazil, Colombia, Mexico and Uruguay) have contributed to earlier studies on physics opportunities at the FCC [2]. More recently, 11 institutes in Latin America (from Brazil, Chile, Colombia and Mexico) have already signed a Memorandum of Understanding to participate in the FCC feasibility study [3].

2.  **LAA-HECAP: a strategic discussion forum for the region**

The Latin American Association for High Energy, Cosmology and Astroparticle Physics (LAA-HECAP) was established in November 2021, following a recommendation from the recent initiative of strategic planning developed by the Latin American Strategy Forum for Research Infrastructures for HECAP (LASF4RI-HECAP). It aims to leverage and amplify the successful and growing dynamics of research in HECAP in Latin America, as demonstrated by the fruitful and long-term organization of the Latin American Symposium on High Energy Physics (Simposio Latino Americano de Física de Altas Energias – SILAFAE) series and LASF4RI-HECAP, that saw the participation of several countries in the region.

Anyone with interest in HECAP can become a member by just filling a form. Currently LAA-HECAP has more than 600 members. The distribution of the members by their career stage is shown in figure 1.

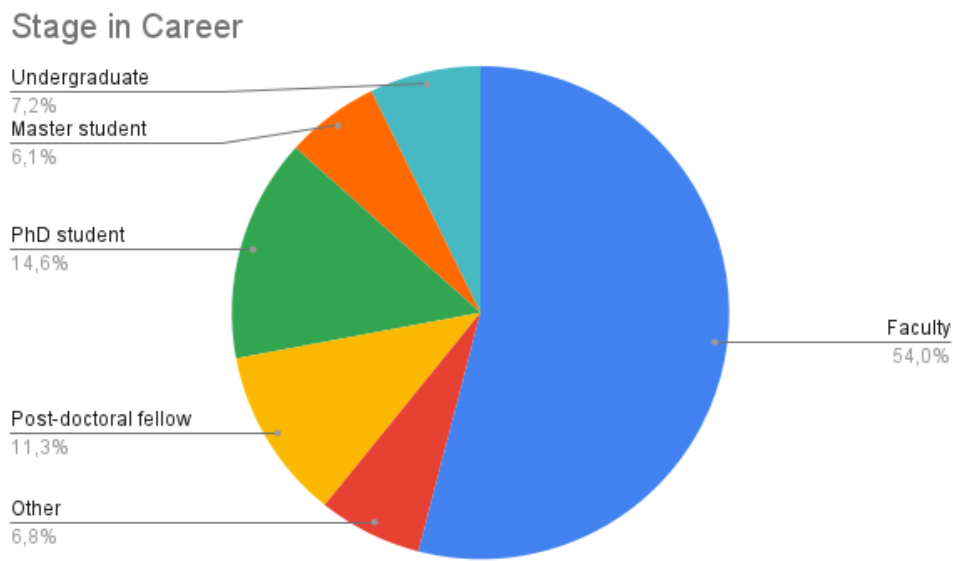

Figure 1: Distribution of LAA-HECAP members by their career stage (as of March 12, 2025).

More than 230 of our members are involved in research in High Energy Experimental Physics, as shown in figure 2.

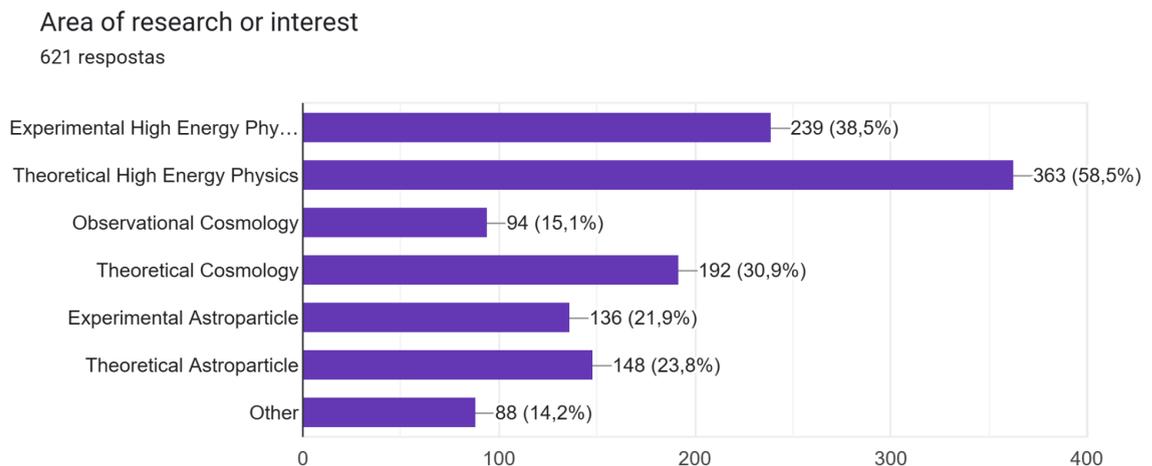

Figure 2: Distribution of LAA-HECAP members by areas of research or interest (as of March 12, 2025).

The overall size of the Latin American community at CERN, both counting researchers affiliated in the region, and elsewhere, makes it relevant to collect their input for the current update of the European Strategy for Particle Physics, and the role of LAA-HECAP, also promoting regional-level strategic discussions, makes it a perfect conduit to do just that. While this cannot be considered a firm position of any institute or funding agency, it reflects a town-hall approach to highlight the prevailing opinions in the region. The process mimicked the ECFA guidelines for inputs from national HEP communities to the European Strategy for Particle Physics [4]. The community input has been gathered through a survey of the LAA-HECAP membership centered around the key open questions for the Strategy Update, and a subsequent town-hall meeting.

3. **Survey methodology**

Following increased interest in the ESPP update in the region, LAA-HECAP charged a committee with collecting, and writing-up, the community input. This committee decided to rely in first instance on a questionnaire (through Google Form) to be distributed to the LAA-HECAP membership. The questions were designed to follow the guidelines suggested by the European Strategy for Particle Physics (ESPP) and closely followed the ECFA guidelines for inputs from national HEP communities to the European Strategy for Particle Physics. The survey was open for 18 days from February 3rd to 21. A Town Hall Meeting took place on March 7, 2025, with an outline of the ESPP update process and a review of the survey results. A lively Q&A session and discussion on the structure of this contribution followed, corroborating the conclusions of the committee.[1]

---

[1] The Google Form can be found at: https://forms.gle/NbdktCGRATjiTmZD6

## 4. Results

In this Section we show the results of the survey. The purpose of the survey was to gather opinions, on an individual and personal basis. We had 77 respondents to the survey. Below we present some of the outcomes.

Almost all respondents are based in Latin America, and 80% are junior or senior faculty. About half are already participating in CERN experiments. The question "In your personal opinion, which should be the preferred next major/flagship collider project for CERN?" had an open text field for the answers. The options of "FCC", "muon collider ", and "CLIC " were mentioned most frequently. Roughly 75% either support (or do not have a strong opinion) that CERN should proceed with its chosen next accelerator irrespective of other developments regarding facilities built elsewhere. It is remarkable that more than 70% of respondents are interested in participating in future activities at CERN, although 80% are not yet involved with any of those (and only 8% mentioned to be currently contributing to in FCC studies).

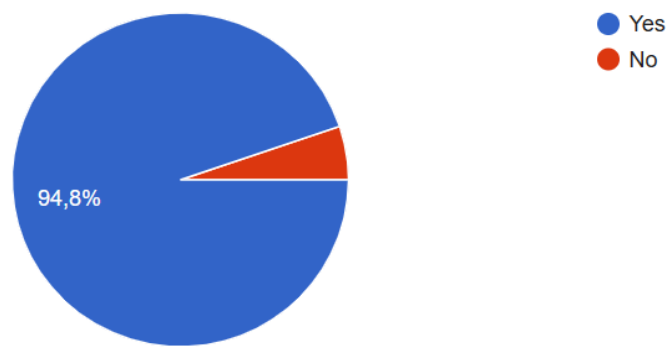

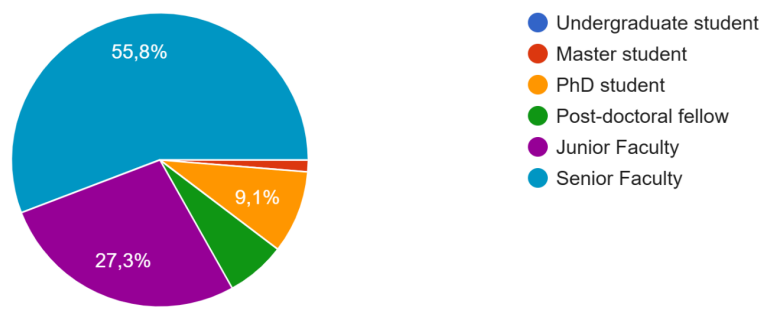

Area of research or interest

77 respostas

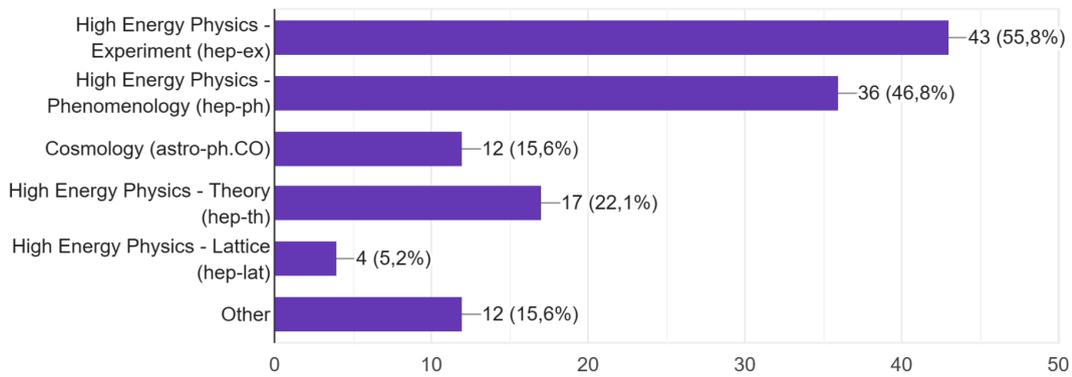

Are you a currently a member of a collaboration at CERN ? If yes, which one?

77 respostas

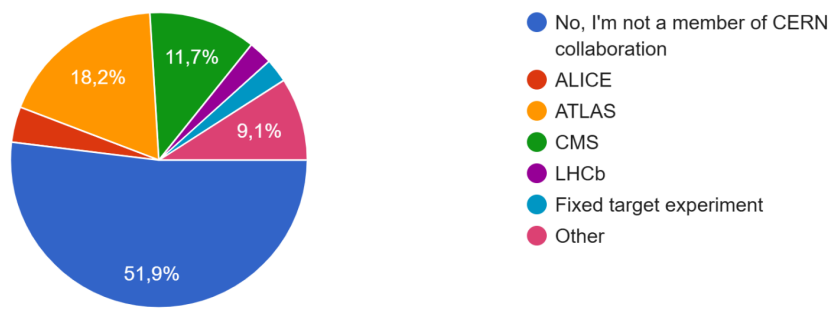

Should CERN/Europe proceed with the preferred option you chose or should alternative options be considered if Japan proceeds with the ILC in a timely way?

77 respostas

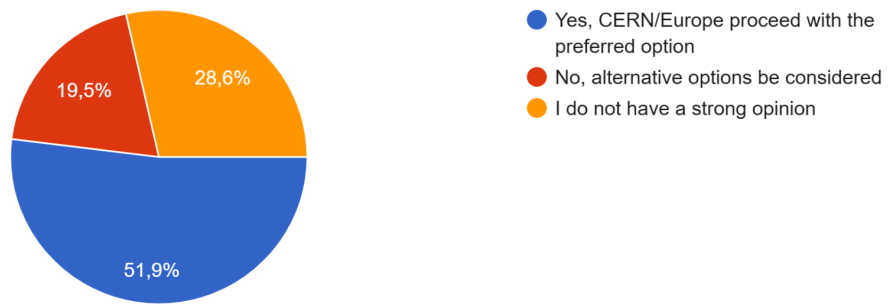

Should CERN/Europe proceed with the preferred option you chose or should alternative options be considered if China proceeds with the CEPC on the announced timescale?

77 respostas

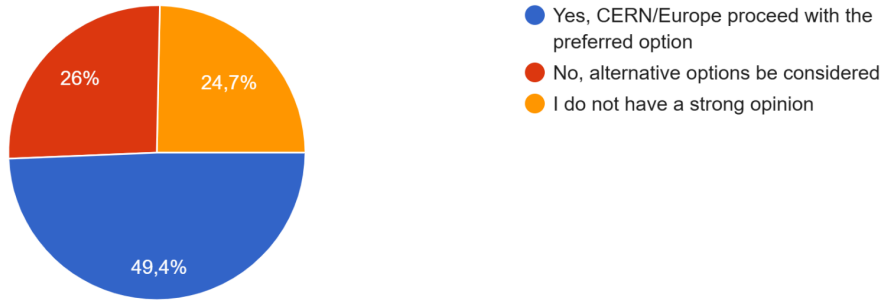

- Yes, CERN/Europe proceed with the preferred option: 49,4%
- No, alternative options be considered: 26%
- I do not have a strong opinion: 24,7%

Should CERN/Europe proceed with the preferred option you chose or should alternative options be considered if the US proceeds with a muon collider?

77 respostas

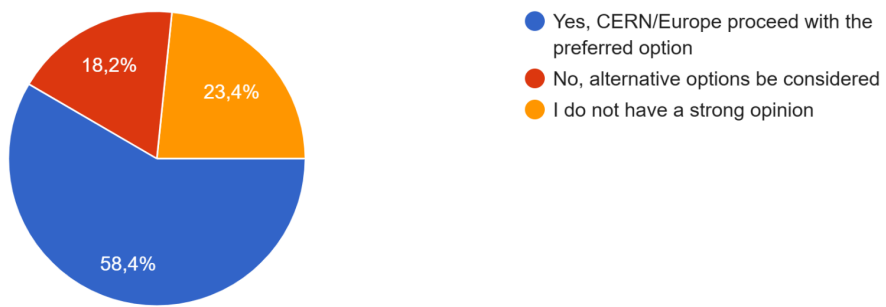

- Yes, CERN/Europe proceed with the preferred option: 58,4%
- No, alternative options be considered: 18,2%
- I do not have a strong opinion: 23,4%

Should CERN/Europe proceed with the preferred option you chose or should alternative options be considered if there are major new (unexpected) results from the HL-LHC or other HEP experiments?

77 respostas

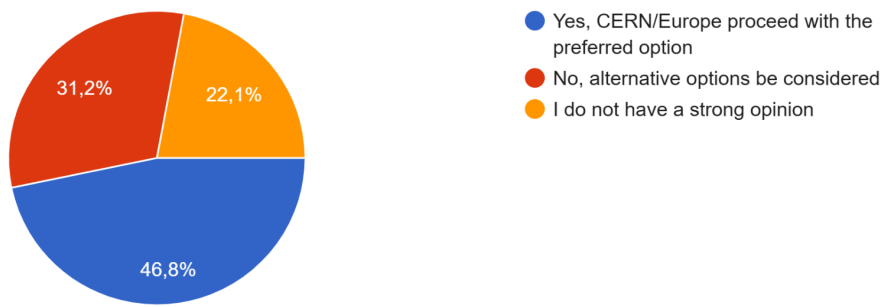

- Yes, CERN/Europe proceed with the preferred option: 46,8%
- No, alternative options be considered: 31,2%
- I do not have a strong opinion: 22,1%

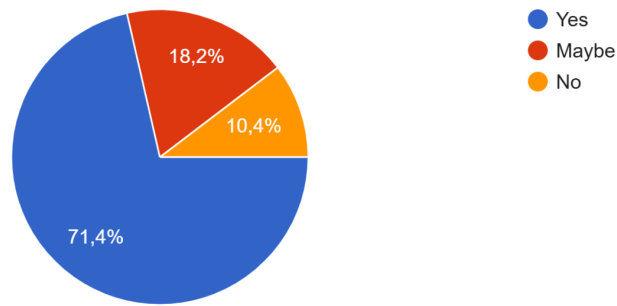

Are you interested in participating in collider physics in the future at CERN?
77 respostas

- Yes 71,4%
- Maybe 18,2%
- No 10,4%

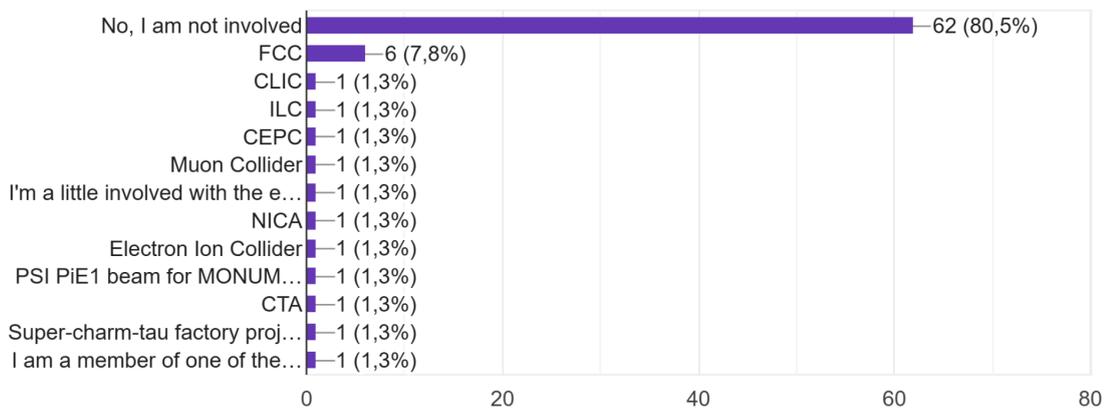

Are you already involved with efforts related to future collider projects?
77 respostas

- No, I am not involved — 62 (80,5%)
- FCC — 6 (7,8%)
- CLIC — 1 (1,3%)
- ILC — 1 (1,3%)
- CEPC — 1 (1,3%)
- Muon Collider — 1 (1,3%)
- I'm a little involved with the e… — 1 (1,3%)
- NICA — 1 (1,3%)
- Electron Ion Collider — 1 (1,3%)
- PSI PiE1 beam for MONUM… — 1 (1,3%)
- CTA — 1 (1,3%)
- Super-charm-tau factory proj… — 1 (1,3%)
- I am a member of one of the… — 1 (1,3%)

5. **Conclusions**

This contribution does not claim to represent the views of any specific country in Latin America, nor the region, and is not prepared within any specific context by national funding agencies. It does present the opinions of a significant number of colleagues in the region, and a significant fraction of the experimental particle physics community. From conversations at the Town Hall meeting, and at various regional conferences, including the most recent SILAFAE event in 2024, we are convinced these opinions represent trends that can be extrapolated to the region as a whole. The most notable of which can be highlighted as:
- There is a considerable interest in the community to engage in a new project at CERN. Remarkably this is from colleagues who are not currently

- participating in CERN's experimental programme, nor are engaged in work on future colliders.
- More than half of the respondents indicated the FCC, possibly with an initial ee phase, to be their preferred option for CERN's next flagship project. Interestingly, approximately 20% of the respondents indicated a muon collider as the machine with the largest physics potential.
- The motivation to participate stands firm in the context of geopolitical competition, suggesting CERN forges ahead with the favorite option, while keeping in consideration possible new physics input from HL-LHC or other discoveries